\documentclass[aps,pra,superscriptaddress,longbibliography,twocolumn]{revtex4-2}

\usepackage{graphicx}
\usepackage{amsmath}
\usepackage{amssymb}
\usepackage{hyperref}
\usepackage[utf8]{inputenc}
\usepackage{mathtools}
\usepackage[english]{babel}
\usepackage{bbm}
\hypersetup{colorlinks=true, linkcolor=blue, citecolor=blue, urlcolor=blue}
\usepackage{xcolor}
\usepackage{layouts}
\usepackage{braket}
\usepackage[normalem]{ulem}
\usepackage{mathrsfs}
\usepackage{dsfont}

\newcommand{\ie}{i.\,e.,\ }

\newcommand{\fref}[1]{\text{Fig.}~\ref{#1}}

\newcommand{\eref}[1]{\text{Eq.}~\eqref{#1}}

\begin{document}
\title{Emergent limit cycles, chaos, and bistability in driven-dissipative atomic arrays}

\author{Victoria Zhang}
\affiliation{Department of Physics, Harvard University, Cambridge, Massachusetts 02138, USA}
\author{Stefan Ostermann}
\affiliation{Department of Physics, Harvard University, Cambridge, Massachusetts 02138, USA}
\author{Oriol Rubies-Bigorda}
\affiliation{Physics Department, Massachusetts Institute of Technology, Cambridge, Massachusetts 02139, USA}
\affiliation{Department of Physics, Harvard University, Cambridge, Massachusetts 02138, USA}
\author{Susanne F. Yelin}
\affiliation{Department of Physics, Harvard University, Cambridge, Massachusetts 02138, USA}

\begin{abstract}
We analyze the driven-dissipative dynamics of subwavelength periodic atomic arrays in free space, where atoms interact via light-induced dipole-dipole interactions. We find that depending on the system parameters, the underlying mean-field model allows four different types of dynamics at late times: a single monostable steady state solution, bistability (where two stable steady state solutions exist), limit cycles and chaotic dynamics.
We provide conditions on the parameters required to realize the different solutions in the thermodynamic limit. In this limit, only the monostable or bistable regime can be accessed for the parameter values accessible via light-induced dipole-dipole interactions. For finite size periodic arrays, however, we find that the mean-field dynamics of the many-body system also exhibit limit cycles and chaotic behavior. Notably, the emergence of chaotic dynamics does not rely on the randomness of an external control parameter but arises solely due to the interplay of coherent drive and dissipation.
\end{abstract}

\maketitle

%==============================================================================================
\section{Introduction} \label{section: intro}
\begin{figure}[t]
    \centering
    \includegraphics[width = 0.95\columnwidth]{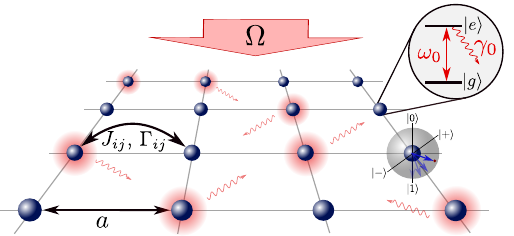}
    \caption{An array of two-level quantum emitters is driven by an external laser with Rabi frequency $\Omega$. When the lattice spacing $a$ is significantly lower than the transition wavelength $\lambda_0=2\pi c/\omega_0$, light-induced dipole-dipole interactions give rise to coherent and dissipative long-range interactions. The underlying driven-dissipative spin model gives rise to complex nonlinear spin dynamics.}
    \label{fig:model}
\end{figure}
Out-of-equilibrium dynamics of long-range interacting many-body quantum systems gives rise to a series of fascinating phenomena. Examples include ergodicity breaking~\cite{turner_weak_2018, serbyn_quantum_2021}, quantum many-body scars~\cite{berry_quantum_1997, ho_periodic_2019, bluvstein_controlling_2021}, time crystals~\cite{zaletel_colloquium_2023, choi_observation_2017, maskara_discrete_2021} and quantum chaos~\cite{jensen_quantum_1992}. While many of these phenomena are usually studied and realized for closed quantum systems, coupling many-body systems to an environment and adding drive can allow enhanced control over the dynamics and the emergence of additional phenomena that have no counterpart in closed systems~\cite{kongkhambut_observation_2022, cabot_quantum_2023, mattes_entangled_2023,ding_ergodicity_2024}. In this work, we study a particular realization of a driven-dissipative system ---  a coherently-driven array of two-level atoms in free space. If the interatomic distance is small enough, light-induced coherent and dissipative long-range interactions give rise to strongly non-linear dynamics in the resultant non-integrable spin model~\cite{Lehmberg_1970_1,Lehmberg_1970_2}.

The recent emergence of technologies that offer enhanced control over the arrangement of individual atoms~\cite{lewenstein_ultracold_2007, bloch_quantum_2012, grier_revolution_2003, barredo_atom-by-atom_2016, endres_atom-by-atom_2016, norcia_microscopic_2018, cooper_alkaline-earth_2018} or quantum emitters~\cite{experiment_quantumdots_1, experiment_quantumdots_2, experiment_NV, experiment_2dmat} in potentially subwavelength geometries requires detailed theoretical analysis of the non-linear dynamics expected for these systems. While many aspects of the transient dynamics were investigated over the past years~\cite{masson_many-body_2020, ferioli_laser-driven_2021, masson_universality_2022, superradiance_cumulants_ours, sierra_dicke_2022, Hanzhen, mok_dicke_2023, rubies-bigorda_dynamic_2023, Oriol_superradiance, Robicheaux_superradiance}, their steady-state properties in the driven-dissipative case were only very recently studied and experimentally analyzed ~\cite{parmee_signatures_2020, ferioli_non-equilibrium_2023, PhysRevResearch.6.023206, ferioli_non-gaussian_2023, Ruostekoski_2024, Agarwal_2024, Goncalves_2024}.

In this work, we go beyond investigating the features of the steady states and perform an in-depth study of the late-time driven-dissipative dynamics of two-dimensional periodic arrays of atoms interacting via light-induced dipole-dipole interactions (see~\fref{fig:model}). Our analysis is based on a mean-field model of the spin-degrees of freedom, which circumvents the complexity of the exponentially growing Hilbert space and allows for investigating large arrays of atoms. We first study configurations where all atoms are permutationally symmetric, which naturally occurs for rings and infinite one- and two-dimensional arrays of emitters,~\ie in the thermodynamic limit. For these configurations, an effective single particle model that describes the dynamics of a single atom in an effective mean-field generated by all the surrounding atoms provides first analytic intuition. We show that the late-time dynamics resulting from this model can in general exhibit monostability (a single steady state solution exists) and bistability (two steady state solutions exist), as well as the emergence of limit cycles (suggesting potential ergodicity breaking) and chaotic dynamics. However, the nature of the effective parameters one obtains for the underlying physical model of dipole-dipole interacting atoms is such that only the mono- and bistability can be accessed in practice. Nonetheless, we demonstrate that all four types of dynamics can be realized for \emph{finite} extended arrays with as little as nine atoms. Transitions between the different regimes can be induced by simply tuning the lattice spacing or the driving strength. This is particularly remarkable for the chaotic regime, which arises solely due to the interplay of dissipation (\ie random quantum jumps) and coherent drive and does not require to add randomness to the system parameters.

%=========================================================================
\section{Model} \label{section: model}
We consider a driven array of $N$ two-level atoms with ground state $\ket{g_i}$ and excited state $\ket{e_i}$ located at positions $\mathbf{r}_i$, as illustrated in~\fref{fig:model}. Applying the Markov approximation and tracing out the field degrees of freedom yields the equations of motion for the atomic operators in the Heisenberg picture~\cite{Lehmberg_1970_1,Lehmberg_1970_2}
\begin{align}
\label{eq: master_equation}
    \frac{d \hat{O}}{d t}=\frac{i}{\hbar}[\hat{H}, \hat{O}]+\mathcal{L}(\hat{O})+\mathcal{F}(\hat{O}).
\end{align}
Assuming the system is driven on resonance, the Hamiltonian in the rotating frame reads
\begin{equation}
    \hat{H} = \hbar \sum_{i,j\neq i}^N J_{ij}\hat{\sigma}_i^+ \hat{\sigma}_j^- + \hbar \frac{\Omega}{2}\sum_{i=1}^N\left(\hat{\sigma}_i^+ + \hat{\sigma}_i^-\right),
\label{eq:Hamiltonian}
\end{equation}
where we have introduced $\hat{\sigma}_i^+ = |e_i \rangle \langle g_i |$ ($\hat{\sigma}_i^- = |g_i \rangle \langle e_i |$) as the raising (lowering) operator for atom $i$. The first term describes coherent exchange interactions between the atoms mediated by the vacuum electromagnetic field, while the second corresponds to a plane wave drive perpendicular to the atomic array with Rabi frequency $\Omega$. The dissipative nature of the system is described by the Lindbladian
\begin{align}
    \mathcal{L}(\hat{O}) &=\sum_{i, j} \frac{\Gamma_{i j}}{2}\left(2 \hat{\sigma}_i^{+} \hat{O} \hat{\sigma}_j^{-}-\hat{\sigma}_i^{+} \hat{\sigma}_j^{-} \hat{O}-\hat{O} \hat{\sigma}_i^{+} \hat{\sigma}_j^{-}\right). \label{eq: dissipative}
\end{align}
The light-induced coherent $J_{ij}$ and dissipative $\Gamma_{ij}$ interactions between atoms $i$ and $j$ are obtained from the Green's tensor for a point dipole in vacuum $\bold{G}$, given in Appendix~\ref{app:Greensfunction}, via
\begin{align}
    J_{i j}-i \Gamma_{i j} / 2 &=-\frac{3 \pi \gamma_0}{\omega_0} \mathbf{d}^{\dagger} \cdot \mathbf{G}\left(\mathbf{r}_{i j}, \omega_0\right) \cdot \mathbf{d},
\end{align}
where $\bold{r}_{ij} = \bold{r}_i- \bold{r}_j$ is the vector connecting atoms $i$ and $j$ and $\bold{d}$ is the transition dipole moment. For the remainder of this work we choose $\textbf{d} = (0, 0, 1)^T$. Here, $\Gamma_{ii} = \gamma_0$ corresponds to the spontaneous decay rate of a single atom in vacuum, and the Lamb shift $J_{ii}$ is included in the definition of the resonance frequency.

\begin{figure}
    \centering
    \includegraphics[width = \columnwidth]{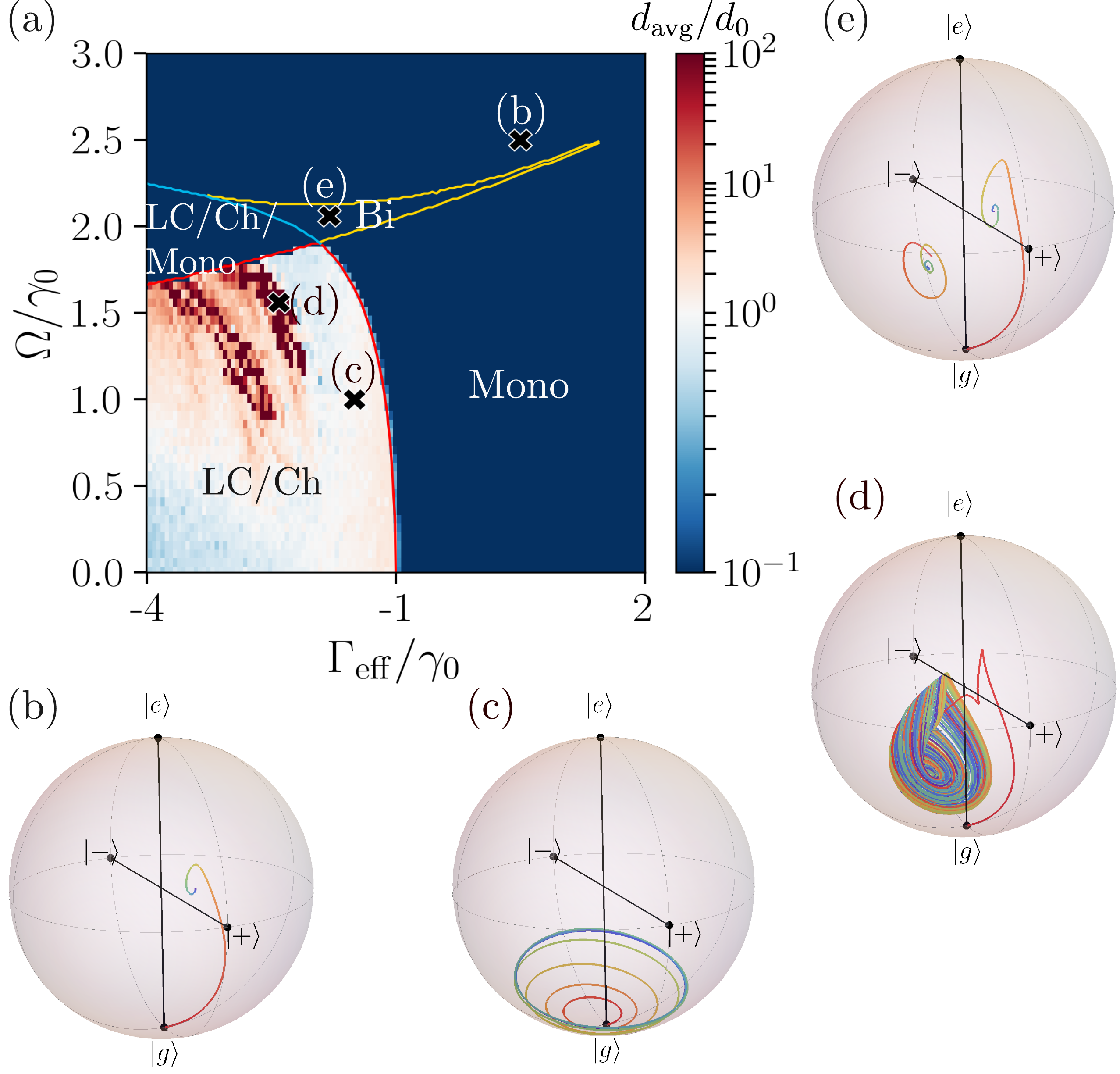}
    \caption{(a) The separation distance $d_\mathrm{avg}$ between two nearby trajectories as a function of the Rabi frequency $\Omega$ and the effective dissipative interaction strength $\Gamma_{\mathrm{eff}}$ for the permutationally symmetric configuration described by ~\eref{eq: thermodynamic mean-field}, with all atoms initialized to their respective ground state. Here, the effective interaction strength is $J_{\mathrm{eff}} = 3\gamma_0$. We also mark the boundaries of the monostable (Mono), bistable (Bi), chaotic (Ch), and limit cycle (LC) regimes in yellow, red, and blue. (b) - (e) Representative phase space trajectories are depicted on the Bloch sphere for the (b) Mono regime where $\Gamma_{\mathrm{eff}}/\gamma_0 = 0.5$ and $\Omega/\gamma_0 = 2.5$, (c) Bi regime where $\Gamma_{\mathrm{eff}}/\gamma_0 = -1.8$ and $\Omega/\gamma_0 = 2.06$, (d) Ch regime where $\Gamma_{\mathrm{eff}}/\gamma_0 = -2.41$ and $\Omega/\gamma_0 = 1.56$, and (e) LC regime where $\Gamma_{\mathrm{eff}}/\gamma_0 = -1.50$ and $\Omega/\gamma_0 = 1.0$, which are marked by crosses in (a). Color represents the passage of time, beginning at red and ending at purple. In (b)-(d), we solve Eq.~\eqref{eq: thermodynamic mean-field} for a system with all atoms initially in the ground state. In (e), we consider two initial conditions to highlight the existence of bistability: first, all atoms initially in the ground state; and second, all atoms initially in the state $ (s_x,s_y,s_z) =(-0.9, 0.1, -0.6) $.}
    \label{fig: thermodynamic phase diagram}
\end{figure}

The last term in~\eref{fig:model}, $\mathcal{F}(\hat{O})$, represents the quantum Langevin noise that arises from vacuum fluctuations~\cite{Lehmberg_1970_1}. Assuming white noise, the expectation value $\langle \mathcal{F}(\hat{O})\rangle$ vanishes. Because we are ultimately interested in the expectation values of atomic operators $\langle \hat{O}\rangle$, we drop $\mathcal{F}(\hat{O})$ from here onward to simplify notation.

Using the relations $\sigma_i^{\pm} = (\sigma_x^i \pm \sigma_y^i)/2$, we obtain the equations of motion for the Pauli operators of the $k^{\rm{th}}$ atom from~\eref{fig:model},
\begin{subequations}
\begin{align}
    \frac{d \hat{\sigma}_x^{k}}{dt} &= - \frac{1}{2} \gamma_0 \sigma_x^{k} + \sum_{i \neq k} J_{ki} \hat{\sigma}_y^{i}\sigma_z^{k} + \frac{1}{2} \sum_{i \neq k} \Gamma_{ki} \hat{\sigma}_x^{i}\hat{\sigma}_z^{k}, \\
    \frac{d \hat{\sigma}_y^{k}}{dt} &= - \Omega \hat{\sigma}_z^{k} - \frac{1}{2} \gamma_0 \hat{\sigma}_y^{k} - \sum_{i \neq k} J_{ki} \hat{\sigma}_x^{i}\hat{\sigma}_z^k + \frac{1}{2} \sum_{i \neq k} \Gamma_{ki} \hat{\sigma}_y^{i}\hat{\sigma}_z^{k}, \\
    \frac{d \hat{\sigma}_z^{k}}{dt} &= \Omega \hat{\sigma}_y^{k} - \gamma_0(\hat{\sigma}_z^{k} + \mathds{1}) - \sum_{i \neq k} J_{ki} (\hat{\sigma}_y^{i}\hat{\sigma}_x^{k} - \hat{\sigma}_x^{i}\hat{\sigma}_y^{k})  \nonumber \\
    & \quad -\frac{1}{2} \sum_{i \neq k} \Gamma_{ki}(\hat{\sigma}_x^{i}\hat{\sigma}_x^{k} + \hat{\sigma}_y^{i}\hat{\sigma}_y^{k}).
\end{align}
\label{eq:pauli}
\end{subequations}
Obtaining the full solution of these equations requires calculating additional equations for the higher-order operators $ \hat{\sigma}^i_y\hat{\sigma}^k_z, \, \hat{\sigma}^i_x\hat{\sigma}^k_z, \,  \hat{\sigma}^i_y\hat{\sigma}^k_x, \, \hat{\sigma}^i_x\hat{\sigma}^k_y, \,  \hat{\sigma}^i_x\hat{\sigma}^k_x, \, \hat{\sigma}^i_y\hat{\sigma}^k_y$, which in turn will again depend on Pauli strings with higher-weight. The number of equations required to describe the system exactly grows exponentially with the number of atoms, making an exact solution of the master equation \eqref{eq: master_equation} unfeasible for large arrays. Hence, in this work, we instead rely on a mean-field approximation of~\eref{eq:pauli} to drastically reduce the number of equations. We apply the mean-field decoupling for the two-point correlators, $\langle \hat{A}\hat{B}\rangle = \langle \hat{A}\rangle \langle\hat{B} \rangle$, and obtain from~\eref{eq:pauli} the equations of motion for the expectation values $s_{x,y,z}^{k}\equiv \langle \sigma_{x,y,z}^{k}\rangle$. They read 
\begin{subequations}
    \begin{align}
        \frac{d s_x^{k}}{dt} &= - \frac{1}{2} \gamma_0 s_x^{k} + \sum_{i \neq k} J_{ki} s_y^{i}s_z^{k} + \frac{1}{2} \sum_{i \neq k} \Gamma_{ki} s_x^{i}s_z^{k}, \\
    \frac{d s_y^{k}}{dt} &= - \Omega s_z^{k} - \frac{1}{2} \gamma_0 s_y^{k} - \sum_{i \neq k} J_{ki} s_x^{i}s_z^k + \frac{1}{2} \sum_{i \neq k} \Gamma_{ki} s_y^{i}s_z^{k}, \\
    \frac{d s_z^{k}}{dt} &= \Omega s_y^{k} - \gamma_0(s_z^{k} + 1) - \sum_{i \neq k} J_{ki} (s_y^{i}s_x^{k} - s_x^{i}s_y^{k}) \nonumber \\
    & \quad - \frac{1}{2} \sum_{i \neq k} \Gamma_{ki}(s_x^{i}s_x^{k} + s_y^{i}s_y^{k}).
    \end{align}
    \label{eq: mean-field}
\end{subequations}
This set of equations is at the core of the analysis presented below.

%=========================================================================
\section{Permutationally Symmetric Configurations} 
\label{section: thermodynamic limit}

We first study the evolution of the system at the mean-field level for permutationally symmetric configurations where all atoms are indistinguishable due to spatial symmetries,~\ie $s_{x, y, z} \equiv s_{x, y, z}^k$ for all $k \in \{1, ..., N\}$. This is the case for periodic arrays in the thermodynamic limit $N \rightarrow \infty$ and for finite geometries such as rings of atoms with perpendicular polarization. Defining the effective coherent and dissipative interaction strengths, $J_\mathrm{eff} = \sum_{i=2}^N J_{1 i}$ and $\Gamma_\mathrm{eff} = \sum_{i=2}^N \Gamma_{1 i}$,~\eref{eq: mean-field} simplifies to three equations  
\begin{subequations}
    \begin{align}
    \frac{d s_x}{dt} &= J_\mathrm{eff} s_y s_z -\frac{1}{2} (\gamma_0 - \Gamma_\mathrm{eff} s_z) s_x \label{eq: thermodynamic mean-field sx},\\
    \frac{d s_y}{dt} &= - \Omega s_z - J_\mathrm{eff} s_x s_z -\frac{1}{2} (\gamma_0 - \Gamma_\mathrm{eff} s_z) s_y \label{eq: thermodynamic mean-field sy}, \\
    \frac{d s_z}{dt} &= \Omega s_y - \gamma_0 (s_z + 1) - \frac{1}{2} \Gamma_\mathrm{eff}(s_x^2 + s_y^2).
    \end{align}
    \label{eq: thermodynamic mean-field}
\end{subequations}
This is an effective single particle model that exactly describes the mean-field dynamics of permutationally symmetric configurations~\cite{Kramer_2016}.

\subsection{Role of effective collective interactions}
\label{subsection: collective shifts}

The dynamics of the system at late times strongly depends on the effective collective interactions $\Gamma_{\rm{eff}}$ and $J_{\rm{eff}}$. To understand this dependency, we characterize the range of possible driven-dissipative dynamics for the permutationally symmetric configuration described by ~\eqref{eq: thermodynamic mean-field} as a function of the effective dissipative interaction $\Gamma_{\rm{eff}}$ and the drive strength $\Omega$ at a fixed effective coherent interaction $J_{\rm{eff}} = 3 \gamma_0$ [see~\fref{fig: thermodynamic phase diagram}(a)]. Note that similar results are observed for other values of $J_{\rm{eff}}$. We combine an analytical and a numerical analysis to characterize the different parameter regimes of~\eref{eq: thermodynamic mean-field}.

\begin{figure}
    \centering
    \includegraphics[width = \columnwidth]{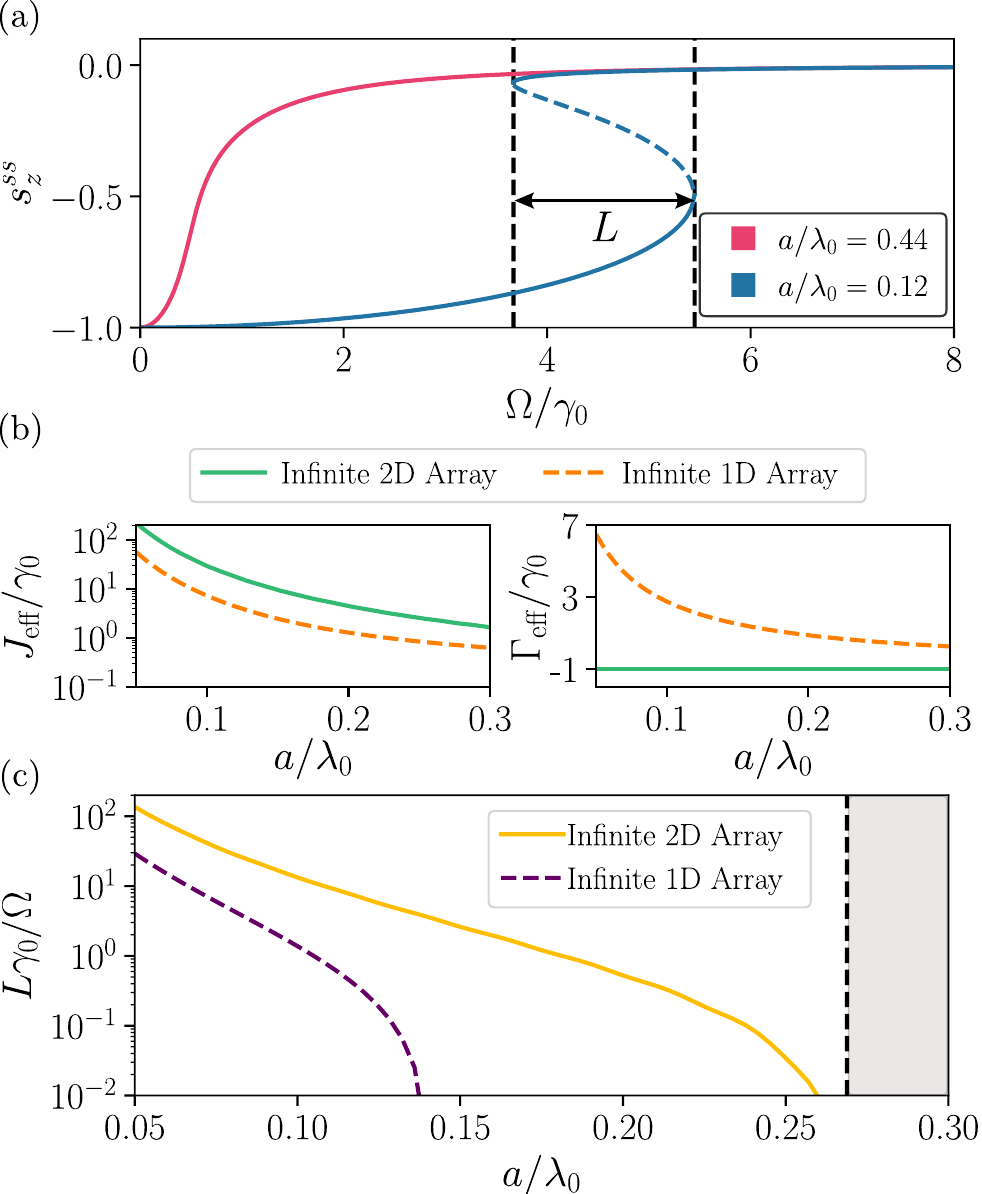}
    \caption{(a) Plot of $s_z^{ss}$ values as a function of driving field strength $\Omega$ for two-dimensional square arrays in the thermodynamic limit, for lattice spacings $a/\lambda_0 = 0.44$ (magenta; monostable) and $a/\lambda_0 = 0.12$ (blue; bistable). Only the physical equilibrium solutions are plotted. The blue dashed line indicates an unstable saddle-focus with a two-dimensional stable manifold. The black vertical dashed lines mark the boundary of the bistable width $L$ for $a/\lambda_0 = 0.12$. (b) The effective coherent interaction strength $J_{\mathrm{eff}}$ and incoherent interaction strength $\Gamma_{\mathrm{eff}}$ as a function of $a$, for both two-dimensional square arrays (solid green) and one-dimensional chain arrays (dashed orange) in the thermodynamic limit. (c) Bistable width $L$ as a function of $a$, for both two-dimensional square arrays (solid yellow) and one-dimensional chain arrays (dashed purple) in the thermodynamic limit. The shaded grey region indicates that bistability cannot occur in two-dimensional square arrays for $a \gtrsim 0.27 \lambda_0$, except for narrow regions around $a / \lambda_0 = 1, \sqrt{2}, ...$ where $J_{\mathrm{eff}}$ and $\Gamma_{\mathrm{eff}}$ diverge.}
    \label{fig: lattice spacing cut}
\end{figure}

Analytically, we characterize the physicality and stability of the equilibrium points of~\eref{eq: thermodynamic mean-field}. Solving for the steady state spin expectation values ($ds^k_\alpha/dt = 0$, $\alpha\in{x,y,z}$) of~\eref{eq: thermodynamic mean-field} yields three equilibrium points $\{s_x^{ss}, s_y^{ss}, s_z^{ss}\}$, which can be nicely distinguished by their $s_z^{ss}$ value [see blue trace in~\fref{fig: lattice spacing cut}(a)]. Mathematically, the steady state solutions may be complex. However, as expectation values must be real, only the purely real fixed points are physically meaningful. We therefore call a steady state solution \emph{physical} if the expectation values of $s_x^{ss}$, $s_y^{ss}$ and $s_z^{ss}$ are real-valued. We then determine the stability by analyzing the eigenvalues of the Jacobian matrix evaluated at each physical equilibrium point and use the following classification terms:

(i) Focus-Node: the Jacobian has one real eigenvalue and a pair of complex-conjugate eigenvalues. All eigenvalues have real parts of the same sign; the focus-node is stable if the sign is negative and unstable if the sign is positive.

(ii) Saddle-Focus: the Jacobian has one real eigenvalue and a pair of complex-conjugate eigenvalues; the sign of the real eigenvalue is opposite that of the real part of the complex-conjugate pair. We say the saddle-focus has a one-dimensional stable manifold if the real eigenvalue is negative; otherwise, it has a two-dimensional stable manifold. A saddle-focus is always unstable along at least one direction in phase space.

Based on the physicality and stability of all equilibrium points (which depend on $J_{\rm{eff}}, \Gamma_{\rm{eff}}$ and $\Omega$), we can draw conclusions on the nature of the resulting dynamics. We say the system is \textit{monostable} if there is only one stable steady state solution. Conversely, we say the system is \textit{bistable} if there are two stable steady state solutions. When the system has a saddle-focus equilibrium point, non-trivial behaviors such as limit cycles and chaotic dynamics can emerge.

Numerically, we characterize the different parameter regimes by evaluating how the distance between two neighboring trajectories evolves as a function of time. More precisely, we initialize all atoms in their ground state and evolve~\eref{eq: thermodynamic mean-field} up to $\gamma_0 t = 2000$ to ensure the orbit has reached its attractor,~\ie the set of states toward which the system tends to evolve. The attractor can either be a point, a cycle, or a chaotic structure. We then choose the state of the system at a random late time close to $\gamma_0 t = 2000$, and apply a small displacement with length $d \ll 1$ to this state. We then propagate the states with and without displacement for an additional $\gamma_0t = 200$ and compute the distance between both orbits as a function of time. We repeat this process for five initial states and eight displacement vectors (see Appendix \ref{app:Lyapunov Exponents} for details), and finally obtain $d_\mathrm{avg}/d_0$, where $d_{\mathrm{avg}}$ is the average distance at the final time and $d_0$ is the average value of $d$. This quantity is inspired by the determination of the Lyapunov exponent~\cite{wolf_lyapunov}, a key quantity used to characterize the dynamics of non-linear systems. While similar conclusions can be reached by analyzing the Lyapunov exponent, $d_\mathrm{avg}/d_0$ is the more robust quantity for the system studied here. 
In particular, it distinguishes the different dynamical behaviors: $d_\mathrm{avg}/d_0$ grows, stagnates and shrinks when the system respectively evolves into chaotic dynamics, a limit cycle, or a stable steady state point.

We identify four regimes with different stability properties [see~\fref{fig: thermodynamic phase diagram}(a)]:

(i) In the monostable (Mono) regime, the only physical equilibrium point is a stable focus-node. All initial conditions evolve to this point at late times. This is confirmed by the distance between neighboring trajectories, $d_\mathrm{avg}/d_0$, which decreases for a system initially in the ground state. A representative steady state trajectory is depicted on the Bloch sphere in~\fref{fig: thermodynamic phase diagram}(b).

(ii) In the bistable (Bi) regime, all three equilibrium points are physical. However, two are stable focus-nodes, while the third is a saddle-focus with a two-dimensional stable manifold. Consequently, all initial conditions reach one of the two stable fixed points, resulting in bistability. This phenomenon is illustrated by the Bloch sphere in~\fref{fig: thermodynamic phase diagram}(e) for two distinct initial conditions. Because of the stable nature of the trajectories, small displacements $d$ result in trajectories evolving into the same stable fixed point and leads to a decrease in $d_\mathrm{avg}/d_0$.

(iii) In the (LC/Ch) regime, the only physical equilibrium point is a saddle-focus with a one-dimensional stable manifold. In this regime, only limit cycle and chaotic behavior are possible. Both types of attractors are distinguished via $d_\mathrm{avg}/d_0$, which remains constant in the former case [white regime in~\fref{fig: thermodynamic phase diagram}(a)] and grows in the latter [red regime in~\fref{fig: thermodynamic phase diagram}(a)]. Representative limit cycle and chaotic trajectories are shown in~\fref{fig: thermodynamic phase diagram}(c) and~\fref{fig: thermodynamic phase diagram}(d), respectively.   

(iv) In the (LC/Ch/Mono) regime, all three equilibrium points are physical: a stable focus-node, a saddle-focus with a one-dimensional stable manifold, and a saddle-focus with a two-dimensional stable manifold. The system can thus evolve to a fixed point, a limit cycle, or a chaotic behavior depending on the initial state and the values of $\Omega$, $\Gamma_{\rm{eff}}$ and $J_{\rm{eff}}$. For the parameters considered in~\fref{fig: thermodynamic phase diagram}(a), the ground state is in the basin of attraction of the fixed point, and $d_\mathrm{avg}/d_0$ consequently decreases.

From~\fref{fig: thermodynamic phase diagram}(a), it becomes apparent that limit cycles and chaos only arise when $\Gamma_{\rm{eff}} < -\gamma_0$. That is, only mono- and bistability are possible for $\Gamma_{\rm{eff}} \geq -\gamma_0$. Intuitively, we can understand this phenomenon from the term $-\frac{1}{2}(\gamma_0 - \Gamma_{\rm{eff}} s_z) \equiv \alpha$ that appears in~\eref{eq: thermodynamic mean-field sx} and~\eref{eq: thermodynamic mean-field sy}. Noting that $s_z^{ss} < 0$, $\alpha$ can be expressed for $s_z < 0$ as
\begin{align}
    \alpha = -\frac{1}{2}(\gamma_0 - \Gamma_{\rm{eff}} s_z) &= -\frac{1}{2} \gamma_0 \left(1+\frac{\Gamma_{\rm{eff}}}{\gamma_0} |s_z|\right).
\end{align}

For $\Gamma_{\rm{eff}}\geq -\gamma_0 $, we obtain $\alpha < 0$. Then, the terms $\alpha s_x$ in~\eref{eq: thermodynamic mean-field sx} and $\alpha s_z$ in~\eref{eq: thermodynamic mean-field sy} are loss terms. For $\Gamma_{\rm{eff}}< -\gamma_0 $, however, we obtain $\alpha > -\gamma_0 / 2$. In particular, one can attain $\alpha > 0$, giving rise to gain in the system and thereby enabling the emergence of limit cycles and chaos.

\begin{figure*}
    \centering
    \includegraphics[width=\textwidth]{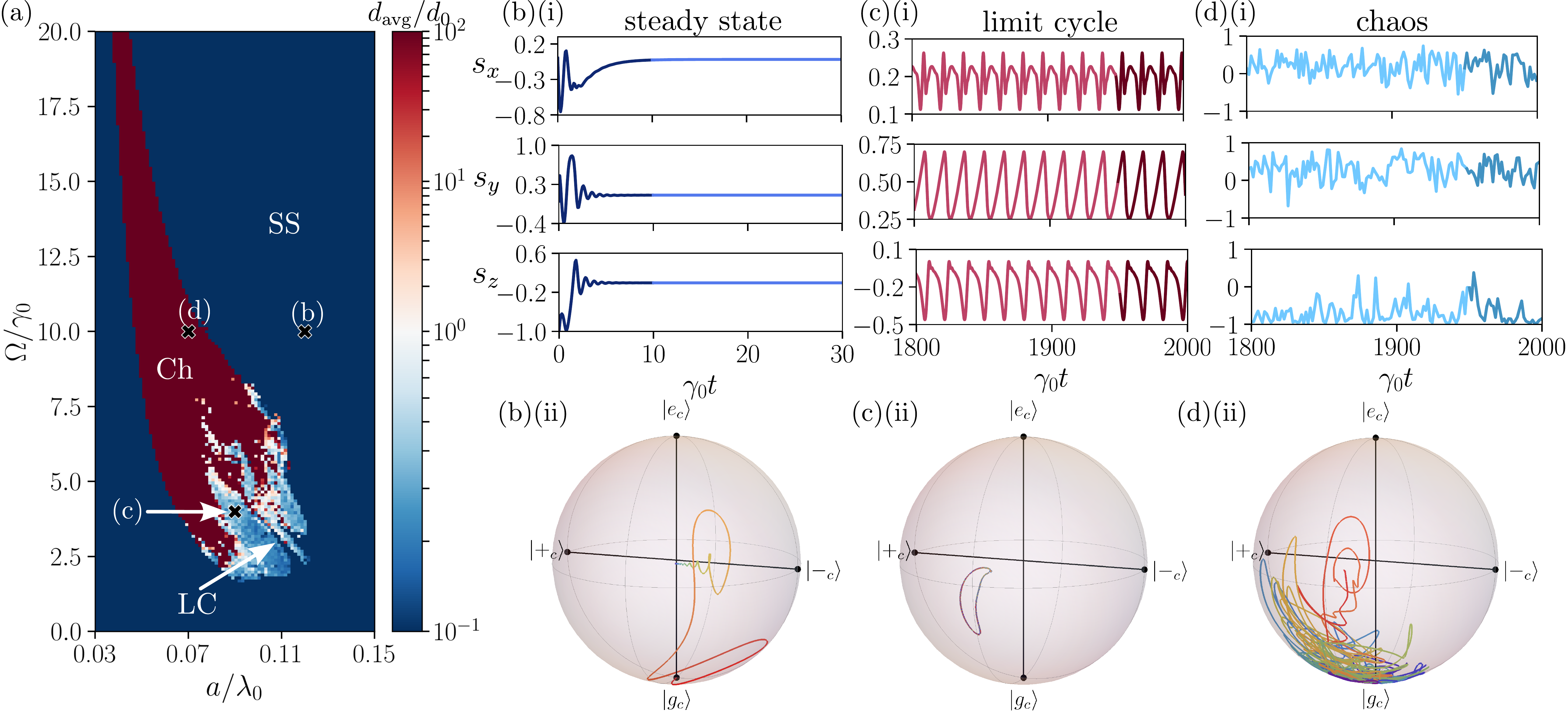}
    \caption{Characterization of the dynamics for a finite size square array with $N=36$ atoms. (a) Separation distance $d_{\mathrm{avg}}$ between two nearby trajectories as a function of Rabi frequency $\Omega$ and lattice spacing $a$. The regimes corresponding to steady state (SS), limit cycles (LC), and chaos (Ch) are labeled. (b) - (d) Time evolution of $s_x$, $s_y$, and $s_z$ of a central atom depicted along with the phase space trajectories on the Bloch sphere. The trajectories are obtained by solving~\eref{eq: mean-field} with all atoms initially in the ground state, and represent the three regimes marked by crosses in (a): (b) steady state dynamics for $a/\lambda_0 = 0.12$ and $\Omega/\gamma_0 = 10$; (c) limit cycle dynamics for $a/\lambda_0 = 0.09$ and $\Omega/\gamma_0 = 4$; and (d) chaotic dynamics for $a/\lambda_0 = 0.07$ and $\Omega/\gamma_0 = 10$. The trajectories in the Bloch sphere are shown in rainbow colors to indicate the passage of time. The time interval chosen corresponds to the darker shaded region in (b)(i)-(d)(i). The states $|e_{c}\rangle$ and $|g_{c}\rangle$ denote the excited and ground state of the central atom, and we additionally define the states $|\pm_{c}\rangle = (|e_{c}\rangle \pm |g_{c}\rangle)/\sqrt{2}$. }
    \label{fig: mean-field phase diagram}
\end{figure*}

\subsection{Dynamics for permutationally symmetric atomic arrays}
For atomic arrays with full permutation symmetry such as rings and periodic lattices with a single atom per unit cell, the minimum effective dissipative interaction is bounded by $\Gamma_{\rm{eff}} = -\gamma_0 $. This follows from the property that the dissipation matrix $\boldsymbol{\Gamma}=\left(\Gamma_{i j}\right)_{i, j=1}^N$ is positive semidefinite~\cite{Nielson_Chuang}. Let $\mathbf{1}_N$ denote the $N \times 1$ column vector of ones. From the definition of a positive semidefinite matrix, it follows
\begin{align}
    \gamma_{\boldsymbol{\Gamma}} &\coloneq \frac{1}{N}\mathbf{1}_N^{\top} \Gamma \mathbf{1}_N \geq 0.
\end{align}
Because all atoms are identical,
\begin{align}
    \sum_{i=1}^N \Gamma_{i, j} &= \gamma_0 + \Gamma_{\rm{eff}}
\end{align}
independently of $j$. Then
\begin{align}
    \gamma_{\boldsymbol{\Gamma}} &= \frac{1}{N} \left(\gamma_0 + \Gamma_{\rm{eff}}\right) \mathbf{1}_N^{\top}\mathbf{1}_N = \gamma_0 + \Gamma_{\rm{eff}} \geq 0.
\end{align}
It readily follows that $\Gamma_{\rm{eff}} \geq -\gamma_0$, which implies that arrays with full permutation symmetry \emph{cannot} exhibit limit cycle and chaotic behaviors regardless of the driving strength $\Omega$. While beyond the scope of the present work, it is worth investigating if engineering alternative dissipation channels or adding incoherent drive can enable the emergence of limit cycles and chaos in such arrays.

Nonetheless, bistability persists in both one-dimensional and two-dimensional arrays within a specific range of lattice spacings. In ~\fref{fig: lattice spacing cut}(a), we plot $s_z^{ss}$ as a function of the drive strength $\Omega$ for a lattice spacing corresponding to monostability (magenta trace) and another corresponding to bistability (blue trace), for two-dimensional square arrays in the thermodynamic limit. We define the bistable width $L$ as the length of the interval $[\Omega_1, \Omega_2]$ that supports bistability. We find $L$ by numerically computing the minimum and maximum $\Omega$ such that all three equilibrium points are physical. The effective couplings for atomic chains and two-dimensional square arrays are plotted in \fref{fig: lattice spacing cut}(b), while the corresponding bistable widths $L$ are shown as a function of spacing $a$ in~\fref{fig: lattice spacing cut}(c).

Bistability for permutationally symmetric geometries occurs for $a \lesssim 0.27 \lambda_0$ for two-dimensional squared arrays and $a \lesssim 0.14 \lambda_0$ for chains. In the case of two-dimensional arrays, additional narrow bistable regimes emerge for spacings where $J_{\mathrm{eff}}$ and $\Gamma_{\mathrm{eff}}$ diverge due to constructive interference, \ie for $a/\lambda_0 = 1$, $a/\lambda_0 = \sqrt{2}$, etc.

%=========================================================================
\section{Finite size arrays} 
As discussed in Section \ref{section: thermodynamic limit}, permutationally symmetric geometries such as rings and infinite one- and two-dimensional arrays do not exhibit limit cycles or chaotic dynamics. This also implies that the dynamics of a minimal square array of four atoms (which is identical to a four-atom ring) is also limited to mono- and bistability at the mean-field level. This raises the question of how increasing the number of particles beyond $N=4$ and breaking the full permutation symmetry affects the system's dynamics.  In this section, we use the full mean-field model described in~\eref{eq: mean-field} to study the late-time behavior of finite two-dimensional atomic square arrays with $N > 4$.

Breaking the full permutation symmetry indeed extends the accessible dynamics beyond just mono- and bistability, and we uncover both limit cycle and chaotic behaviors in lattices containing as little as nine atoms (see Appendix \ref{app:N dependency}). For the set of equations~\eqref{eq:  mean-field}, an analytical stability analysis is intractable. We rely on the numerical stability analysis based on the system's response to a perturbation once it reaches its attractor. In~\fref{fig: mean-field phase diagram}(a), we show the stability analysis for a square lattice with $N=36$ atoms initially in the ground state. Exemplary phase space trajectories on the Bloch sphere corresponding to steady state, limit cycle, and chaotic dynamics are shown respectively in~\fref{fig: mean-field phase diagram}(b) through ~\fref{fig: mean-field phase diagram}(d). These dynamics can be nicely distinguished by the growth of the separation distance $d_\mathrm{avg}/d_0$ between two nearby trajectories (for details see Appendix \ref{app:Lyapunov Exponents}). Note that similar results can be obtained for other atom numbers (see \ref{app:N dependency}). The size of the regime exhibiting non-trivial limit cycle or chaotic dynamics changes with system size. Based on the findings presented in section \ref{section: thermodynamic limit}, this regime is expected to vanish for $N\rightarrow \infty$ once the lattice approaches the thermodynamic limit.

\begin{figure}
    \centering
    \includegraphics[width = \columnwidth]{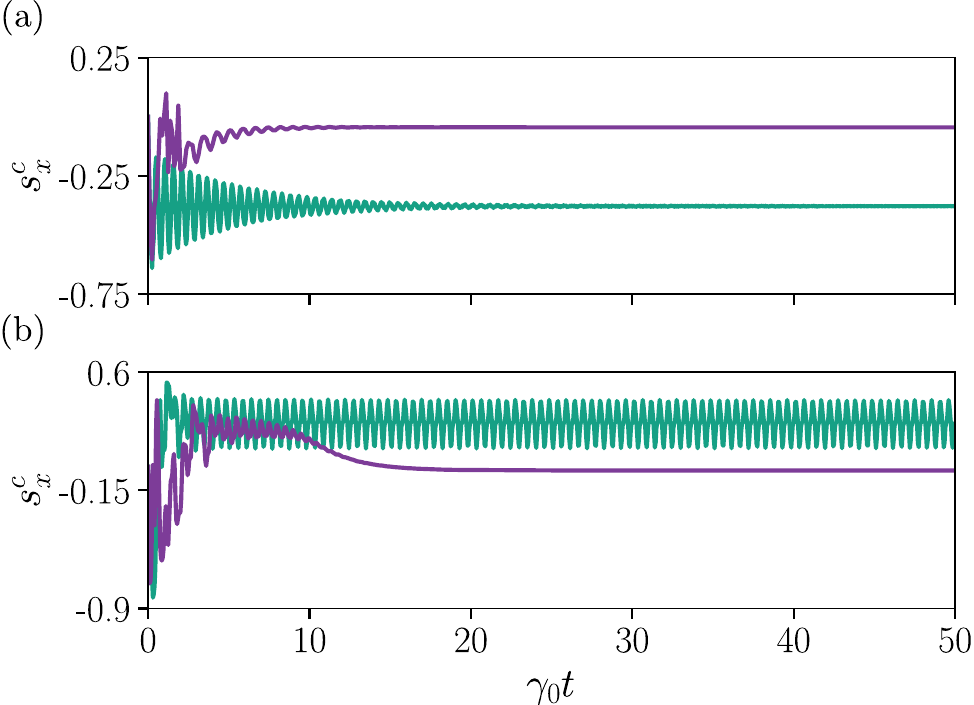}
    \caption{Time evolution of the $s_x$ spin component for the central atom of a nine square atom array, for (a) a bistable example obtained for a lattice spacing $a/\lambda_0 = 0.05$ and driving field strength $\Omega/\gamma_0 = 7.778$, and (b) a dual behavior example with $a/\lambda_0 = 0.0667$ and $\Omega/\gamma_0 = 8.889$. Each example solves~\eref{eq: mean-field} for two initial states: all atoms initialized to their respective ground state (green) and all atoms initialized to their respective excited state (purple). In (a), the green and purple lines correspond to two distinct stable steady states. In (b), the purple line corresponds to a stable steady state, while the green line corresponds to a limit cycle.}
    \label{fig: bistability/dual behavior}
\end{figure}

Beyond limit cycles and chaos, we also observe that bistability and dual behaviors persist at the mean field level for finite sized arrays. \fref{fig: bistability/dual behavior}(a) illustrates an example of bistability within a nine atom square array, where different initial states lead the central atom to evolve towards distinct steady states. \fref{fig: bistability/dual behavior}(b) illustrates a dual behavior example, where the central atom can evolve towards a steady state or a limit cycle depending on the initial state. Note that we have not ruled out the possibility of other dual and more complex behaviors, such as multistability. While such an analysis warrants further study, it goes beyond the scope of the present work. 

Finally, it is worth noting that the amplitude of limit cycle and chaotic dynamics is significantly reduced for the averaged spin components (e.g. $s_z =\frac{1}{N} \sum_k s_z^k$). Although each atom individually exhibits either limit cycles or chaotic behavior, they oscillate or fluctuate out of phase and centered around different average values. Consequently, the average spin expectation value exhibits much smaller oscillations in time.

\section{Conclusions and Outlook}
We have analyzed the mean-field dynamics of a driven dissipative array of atoms in a two-dimensional periodic lattice. Based on an effective single particle mean-field model that captures permutationally symmetric configurations, we determined the possible non-trivial dynamics that can occur in such a system by tuning the effective parameters. Interestingly we find that only mono- and bistability persist for the parameter space accessible for dipole-dipole coupled atom arrays with permutationally symmetric geometry. However, this limitation vanishes for the case of finite system sizes, where we find limit cycles, chaotic dynamics, and bistability in addition to the trivial monostable steady state solution. The different types of dynamics can be accessed by simply changing the spacing of the lattice or the strength of the driving field.

While a treatment of the full quantum model is intractable for the system sizes required to attain non-ergodic dynamics over a wide range of parameters, our findings motivate further studies including quantum correlations. Because theoretical treatments in this case will always be limited to approximate numerical methods, this also renders an exciting avenue for experiments. Our findings unveil the fascinating many-body dynamics that the new generation of subwavelength optical lattice or tweezer experiments can uncover by combining dissipation and drive. In the full quantum case, the chaotic dynamics could also provide an intriguing route towards fast scrambling or the generation of spin squeezed states in driven dissipative light-matter systems.

The observed non-ergodic behavior in dissipative finite-size systems with long-range interactions also provides an ideal test bed to analyze the role of dissipation and the interaction range for ergodicity breaking in open quantum systems~\cite{ding_ergodicity_2024, jiao_observation_2024}. Hence, our work paves the way for future explorations of many-body quantum dynamics in open quantum systems and has potential implications for technological advancements in areas such as quantum sensing, information processing and quantum simulation.

\begin{acknowledgments}
The authors would like to thank Ana-Maria Rey and Na Li for useful discussions. V.Z. was supported by the Harvard College Research Program and the Herchel Smith Undergraduate Science Research Program. S.O. is supported by a postdoctoral fellowship of the Max Planck Harvard Research Center for Quantum Optics. O.R.B. acknowledges support from Fundación Mauricio y Carlota Botton and from Fundació Bancaria “la Caixa” (LCF/BQ/AA18/11680093). S.F.Y would like to acknowledge funding from NSF through the CUA PFC PHY-2317134, the Q-SEnSE QLCI OMA-2016244 and PHY-2207972. The numerical results were obtained using the \textsc{Quantumoptics.jl} package~\cite{kramer_quantumopticsjl_2018} partially using Harvard University’s FAS Research Computing infrastructure.
\end{acknowledgments}

%==============================================================================================
\vspace{0.5cm}

\onecolumngrid
\appendix
%==============================================================================================
\section{Green's function}\label{app:Greensfunction}
The Green's function for a point dipole that determines the interaction strength $J_{ij}$ and the collective dissipation $\Gamma_{ij}$ in~\eref{eq:Hamiltonian} and~\eref{eq: dissipative} can be written in Cartesian coordinates as~\cite{GreensFunction_Chew,GreensFunction_novotny_hecht_2006}
\begin{align}
G_{\alpha \beta}(\mathbf{r},\omega) &= \frac{e^{i k r}}{4\pi r} \left[ \left( 1 + \frac{i}{kr} - \frac{1}{(kr)^2} \right) \delta_{\alpha\beta} \right. \nonumber \\
  &+ \left. \left(-1 - \frac{3i}{kr} + \frac{3}{(kr)^2} \right) \frac{r_\alpha r_\beta}{r^2} \right] + \frac{\delta_{\alpha \beta} \delta^{(3)}(\mathbf{r})}{3k^2},
\end{align}
where $k=\omega_0/c$, $r=|\mathbf{r}|$, and $\alpha,\beta=x,y,z$.

\section{Determining the separation distance between two nearby trajectories}\label{app:Lyapunov Exponents}
For the dynamical system of the form 
\begin{align}
\dot{\mathbf{x}}&=\mathbf{f}(\mathbf{x}, \gamma_0 t),\label{eq: model}
\end{align}
the maximal Lyapunov exponent $\lambda$ indicates the chaotic or regular nature of orbits and characterizes the rate of separation of infinitely close trajectories~\cite{Skokos2010}. It is known that in a three-dimensional system, $\lambda > 0$ for a chaotic attractor; $\lambda = 0$ for a limit cycle;  and $\lambda < 0$ for a steady state solution~\cite{wolf_lyapunov}. We estimate $\lambda$ by monitoring the rate of the change of the distance $|\Delta \tilde{\mathbf{x}}(\gamma_0 t)|$ between a pair of initially close trajectories, where~\cite{Skokos2010}
\begin{align}
    |\Delta \tilde{\mathbf{x}}(\gamma_0 t)| &\approx|\Delta \tilde{\mathbf{x}}(0)| e^{\lambda \gamma_0 t}. 
\end{align}
In particular, for the thermodynamic limit, we implement the following algorithm given an effective dissipative interaction strength $\Gamma_{\rm{eff}}$ and driving field strength $\Omega$:

\begin{figure*}
    \centering
    \includegraphics[width=\textwidth]{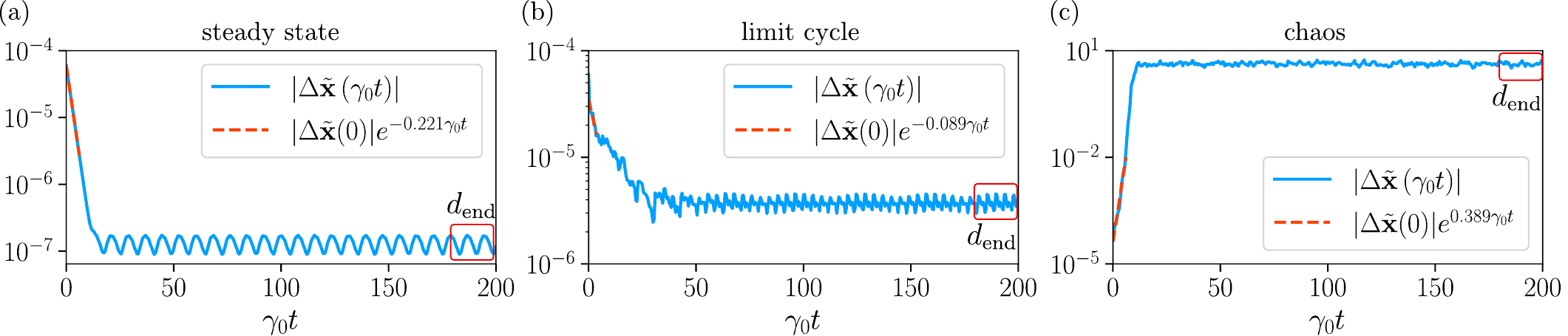}
    \caption{Evolution of the separation distance $|\Delta \tilde{\mathbf{x}}(\gamma_0 t)|$ between two neighboring orbits (solid blue) and the linear fit (dashed red) for a thirty-six atom square array, for (a) a steady state example, where $a/\lambda_0 = 0.10$ and $\Omega/\gamma_0 = 10$, (b) a limit cycle example, where $a/\lambda_0 = 0.09$ and $\Omega/\gamma_0 = 6$, and (c) a chaos example, where $a/\lambda_0 = 0.07$ and $\Omega/\gamma_0 = 10$. For each example, we first solve~\eref{eq: mean-field} up to $\gamma_0 t = 2000$ using initial condition (0, 0, -1) for all atoms. We choose $(s_x^{1}, .., s_x^{36},s_y^{1}, .., s_y^{36}, s_z^{1}, .., s_z^{36})$ evaluated at $\gamma_0 t = 1600$ as the point on the attractor and $(s_x^{1} + \epsilon, .., s_x^{36}+ \epsilon,s_y^{1}, .., s_y^{36}, s_z^{1}, .., s_z^{36})$ as the nearby point, where $\epsilon = 10^{-5}$. We advance each orbit up to $\gamma_0 t = 200$, compute $|\Delta \tilde{\mathbf{x}}(\gamma_0 t)|$, and fit the distance as described in Appendix \ref{app:Lyapunov Exponents} to estimate the maximal Lyapunov exponent $\lambda$. We find $\lambda = -0.221$ for (a), $\lambda = -0.080$ for (b), and $\lambda = 0.389$ for (c). We also mark $d_{\mathrm{end}}$ as the average final separation distance from $\gamma_0 t = 180$ to $\gamma_0 t = 200$.} 
    \label{fig: distance growth}
\end{figure*}

\begin{enumerate}
    \item Solve~\eref{eq: thermodynamic mean-field} [which takes the form of~\eref{eq: model}] up to $\gamma_0 t = 2000$ with all atoms initially in their respective ground state.
    \item Choose $\mathbf{x}_0 = \mathbf{x}(\gamma_0 t = 1600)$ as a point on the attractor.
    \item For each separation vector $\mathbf{d}$ in the list
    \begin{align}
        [(\epsilon, 0, 0), (-\epsilon, 0, 0), (0, \epsilon, 0), (0, -\epsilon, 0), (0, 0, \epsilon), (0, 0, -\epsilon), (\epsilon, \epsilon, \epsilon), (-\epsilon, -\epsilon, -\epsilon)], \label{eq: separation vectors}
    \end{align}
    choose $\mathbf{x}_0+\mathbf{d}$ as the nearby point, where $\epsilon = 10^{-5}.$ These eight separation vectors are chosen to encompass the three dimensional space of the Bloch sphere.
    \item Advance each orbit up to $\gamma_0 t = 200$ for both $\mathbf{x}_0$ and $\mathbf{x}_0+\mathbf{d}$ as initial conditions. 
    \item Calculate the separation distance $\left|\Delta \tilde{\mathbf{x}}\left(\gamma_0 t\right)\right|$ between the two nearby orbits, and compute the average final separation distance $d_{\mathrm{end}}$ from $\gamma_0 t= 180$ to $\gamma_0 t = 200$.
    \item Define $t_{\mathrm{fit}}$ as the first instance in which $|\Delta \tilde{\mathbf{x}}(\gamma_0 t_{\mathrm{fit}})| = \frac{1}{2} |d_{\mathrm{end}} - d|$. Fit an exponential curve up to $t_{\mathrm{fit}}$ to estimate the Lyapunov exponent $\lambda$. 
    \item Repeat steps 3-6 for $\mathbf{x}_0 = \mathbf{x}(\gamma_0 t = 1700)$, $\mathbf{x}_0 = \mathbf{x}(\gamma_0 t = 1800)$, $\mathbf{x}_0 = \mathbf{x}(\gamma_0 t = 1900)$, $\mathbf{x}_0 = \mathbf{x}(\gamma_0 t = 2000)$ for a total of forty trials. Average over all trials to obtain an average estimate of $\lambda$ and an average value $d_{\rm{avg}}$ for $d_{\mathrm{end}}$.
\end{enumerate}

For the finite mean field model in~\eref{eq: thermodynamic mean-field}, the procedure is analogous, except that the separation vector $\mathbf{d}$ is applied uniformly to the corresponding coordinate of every particle. For instance, the separation vector $\mathbf{d} = (\epsilon, 0,0)$ becomes 
\begin{align}
    \mathbf{d} &= (\underbrace{\epsilon, \epsilon, \ldots, \epsilon}_{N \text { times }}, \underbrace{0,0, \ldots, 0}_{N \text { times }}, \underbrace{0,0, \ldots, 0}_{N \text { times }}),
\end{align}
and likewise for other separation vectors in~\eref{eq: separation vectors}. Here, we use the notation $\mathbf{x} = (s_x^1, .., s_x^{N}, s_y^1, .., s_y^{N}, s_z^1, .., s_z^{N}).$ 

However, for our models in~\eref{eq: mean-field} and~\eref{eq: thermodynamic mean-field}, we find that $\lambda$ is unable to distinguish between steady state and limit cycles. In particular, while $\lambda > 0$ for chaotic behavior (see~\fref{fig: distance growth}(c)) as desired, $\lambda < 0$ for both steady state and limit cycles (see~\fref{fig: distance growth}(a) and~\fref{fig: distance growth}(b)). This arises due to the sensitivity of the fitting process in Step 6 to the choice of $t_{\rm{fit}}$. Nonetheless, the separation distance itself, rather than the rate of separation, nicely characterizes the three different dynamical behaviors, as shown in~\fref{fig: thermodynamic phase diagram}(a) and~\fref{fig: mean-field phase diagram}(a). Here, $d_0 = (2+\sqrt{3})/3 \epsilon$ is the average initial separation distance for the thermodynamic limit, and $d_0 = \sqrt{N} \left(\frac{2+\sqrt{3}}{3}\right) \epsilon$ for the finite limit case.

We define $d_{\rm{avg}}/d_0 \leq 10^{-1}$ as steady state, $d_{\rm{avg}}/d_0 > 10^{0}$ as chaos, and intermediate values as limit cycles. 

\begin{figure*}
    \centering
    \includegraphics[width=\textwidth]{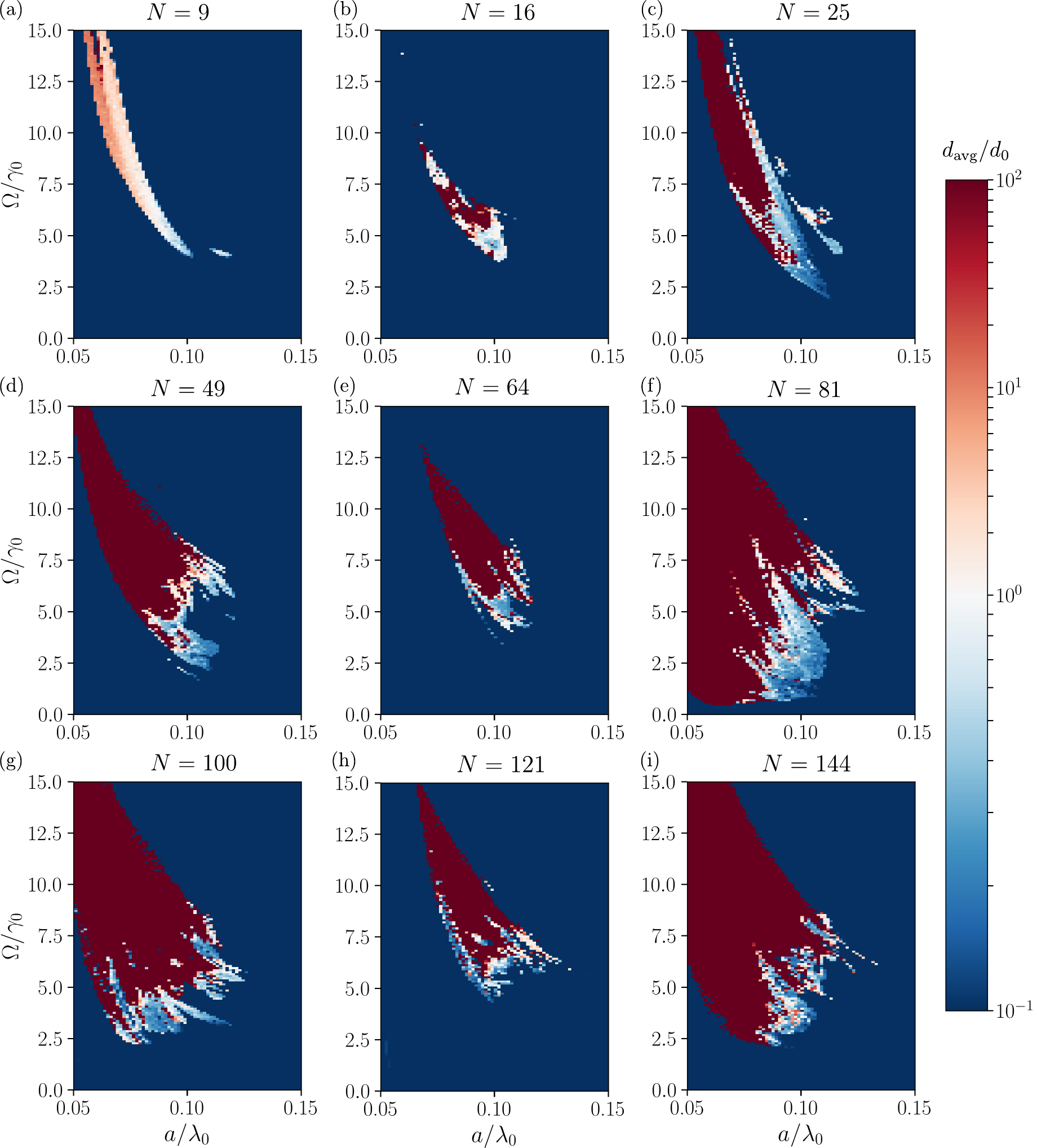}
    \caption{Separation distance $d_{\rm{avg}}$ between two nearby trajectories as a function of Rabi frequency $\Omega$ and lattice spacing $a$ for two-dimensional square arrays,  across varying system sizes from $N = 3^2$ to $N = 12^2$. The dynamics are computed for arrays with all atoms initially in the ground state.}
    \label{fig: N dependency}
\end{figure*}
 
\section{Dependency of non-steady state dynamics on $N$}\label{app:N dependency}
In the main text, we noted that limit cycles and chaotic behavior arise in two-dimensional squared arrays with as little as nine atoms. In~\fref{fig: N dependency}, we show the growth of the separation distance $d_\mathrm{avg}$ between two nearby trajectories as a function of $N$ from $N = 3^2$ up to $N=12^2$ with all atoms initially in the ground state. These finite system sizes are currently available in state-of-the-art experimental setups. We note that there appears to be no clear relationship between $N$ and the size of the nontrivial limit cycle and chaotic regimes.

\twocolumngrid
\bibliography{references}
\end{document}